\documentclass[12pt]{article}
\pdfoutput=1  %needed for JHEP at least?
\usepackage{hyperref}
\usepackage[pdftex]{graphicx}
\usepackage{amsmath,amssymb}
\usepackage{cancel}
\usepackage{appendix}
\usepackage{mathtools}
\usepackage{xcolor}
\usepackage[mathscr]{eucal}
\begin{document}
\begin{titlepage}

\begin{center}
\Large \bf \textbf{de Sitter Holography and\\ Carrollian Brane Theories}
\end{center}

\begin{center}
Andrés Argandoña, Alberto G\"uijosa and Sergio Patiño-López%$^{\ast}$

\vspace{0.2cm}
%$^{\ast}\,$
Departamento de F\'{\i}sica de Altas Energ\'{\i}as, Instituto de Ciencias Nucleares, \\
Universidad Nacional Aut\'onoma de M\'exico,
\\ Apartado Postal 70-543, CdMx 04510, M\'exico\\
 \vspace{0.2cm}
\vspace{0.2cm}
{\tt andres.argandona@correo.nucleares.unam.mx,
alberto@nucleares.unam.mx,fis.selp16@gmail.com}
\vspace{0.2cm}
\end{center}

\begin{center}
{\bf Abstract}
\end{center}
\noindent
It was discovered in recent months that the anti-de Sitter (AdS) backgrounds involved in all familiar top-down examples of AdS/CFT duality follow from applying a transverse \emph{nonrelativistic} brane limit to string/M theory on (asymptotically) flat spacetime. \hfill In \hfill this \hfill note \hfill we \hfill show \hfill that \hfill an \hfill exactly \hfill analogous \hfill statement \hfill holds \hfill for \\ \emph{de Sitter} (dS) backgrounds relevant to particular instances of dS/CFT duality, which are obtained via a longitudinal \emph{Carrollian} brane limit. This statement makes direct contact with the holographic duality inferred by Hull via temporal T-duality. 
 
 \vspace{0.4cm}
\smallskip
\end{titlepage}

\tableofcontents

\section{Introduction and Summary}
\label{introsec}

Surprisingly, it was discovered in recent months \cite{Blair:2024aqz,Guijosa:2025mwh} that each of the brane-based instances of \emph{relativistic} holography, including familiar examples such as
\cite{Maldacena:1997re,Itzhaki:1998dd,Aharony:1998ub,Klebanov:1998hh,Seiberg:1999xz,Maldacena:1999mh}, is in fact a statement within one of the \emph{nonrelativistic} (NR) brane theories that were formulated a quarter century ago \cite{Danielsson:2000gi,Gomis:2000bd,Danielsson:2000mu}  as interesting simplified corners of string/M theory. Key to this discovery was the insight that black brane backgrounds in NR brane theories typically interpolate between an inner relativistic core and an outer nonrelativistic asymptopia \cite{Avila:2023aey} (see also \cite{Danielsson:2000mu,Guijosa:2023qym}). A further, remarkable realization \cite{Blair:2024aqz} was that the gravitational backgrounds in brane-based holography are exactly of this asymptotically NR form, and that the usual near-horizon limit through which they are derived coincides in each case with the limit that defines a corresponding NR brane theory. Taken together, these insights led to the conclusion \cite{Guijosa:2025mwh} that relativistic holography is precisely equivalent to the statement of open-string/closed-string (or more generally, brane/black-brane) duality within that nonrelativistic framework.\footnote{See the bibliographies of \cite{Blair:2024aqz,Guijosa:2025mwh} for earlier related references. Subsequent developments include \cite{Harmark:2025ikv,Bergshoeff:2025grj,Lescano:2025ixp,Blair:2025prd,Blair:2025ewa,Banerjee:2025gyh,Bergshoeff:2025uut,Lescano:2025yio}.}

To appreciate this in more detail, consider the celebrated D3-brane-based duality, which involves Type IIB string theory on the AdS$_5\times$S$^5$ geometry
\begin{equation}
\begin{aligned}
ds^2&=\frac{r^2}{L^2}\left(-dx_0^2+\ldots+dx_3^2\right)+\frac{L^2}{r^2}dr^2+L^2 d\Omega_5^2
\\
&=\frac{r^2}{L^2}\left(-dx_0^2+\ldots+dx_3^2\right)+\frac{L^2}{r^2}\left(dx_4^2+\ldots+dx_9^2\right)~,
\end{aligned}
\label{adsmetric}
\end{equation}
where $r^2\equiv x_4^2+\ldots+x_9^2$. The second line makes it clear that the asymptotic behavior as $r\to\infty$ agrees \cite{Blair:2024aqz} with the scaling that defines the NR D3-brane theory\footnote{This theory and its generalizations to other values of $p$ were originally called Wrapped D$p$-brane theories (WD$p$) in \cite{Danielsson:2000gi} and (with a  narrower perspective) Galilean D$p$-brane theories (GD$p$) in \cite{Gomis:2000bd}. We adopt here the same perspective as in the former reference, but use the NR monicker that was proposed in the latter reference (for the NRF1 case dual to NRD1), because it has become standard. In the recent work \cite{Blair:2024aqz}, NRD$p$ theories were discussed under the alternative name {\bf M}$p${\bf T}.} \cite{Danielsson:2000gi,Gomis:2000bd}, namely
\begin{equation}
ds^2=\omega\, 
\left(-dx_0^2+\ldots+dx_3^2\right)
+\omega^{-1}
\left(dx_4^2+\ldots+dx_9^2\right)
\label{nrd3metricscaling}
\end{equation}
with $\omega \to \infty$, holding the string length and string coupling fixed. 
This scaling tears apart the Lorentzian metric into two separate pieces, one longitudinal and the other transverse to the D3-branes, giving rise to a 3-brane generalization of Newton-Cartan geometry \cite{Andringa:2012uz,Kluson:2017abm,Pereniguez:2019eoq,Blair:2021waq,Ebert:2021mfu,Novosad:2021tlq,Bergshoeff:2023rkk,Ebert:2023hba,Bergshoeff:2024ipq}. By design, the limit flattens out the lightcone in the transverse directions, implying that physics in that sector becomes nonrelativistic. On the other hand, the lightcone along the longitudinal directions is kept at 45 degrees, so physics in that subspace remains relativistic.  

With this reinterpretation, one comes to understand \cite{Blair:2024aqz,Guijosa:2025mwh} that the near-horizon limit \cite{Maldacena:1997re} that yields (\ref{adsmetric}) by dropping the `1' from the harmonic function $H(r)=1+L^4/r^4$ in the parent extremal D3 black brane \cite{Horowitz:1991cd} does not in fact eliminate the asymptotically flat Minkowskian region. Instead, it transmutes it into an asymptotically flat Newton-Cartan region. The AdS$_5\times$S$^5$ geometry is thus recognized \cite{Guijosa:2025mwh} as the extremal D3 black brane of NRD3 theory, which interpolates between an inner  region ($r\lesssim L$) that encodes the relativistic physics of the D3-brane stack and an outer nonrelativistic region ($r\gg L$) that is situated away from the stack. A particularly interesting corollary \cite{Guijosa:2025mwh} is that the bare Newton-Cartan geometry is the preexisting foundation on top of which the Lorentzian spacetime is built by the pattern of entanglement of the D3-brane degrees of freedom. Parallel statements apply for D$p$-branes with $p\neq 3$, as well as for other types of branes and brane intersections \cite{Blair:2024aqz,Guijosa:2025mwh}. 

Knowing that a nonrelativistic limit of asymptotically flat string theory yields \emph{anti}-de Sitter spacetime in the settings relevant to the AdS/CFT correspondence \cite{Maldacena:1997re}, it is natural to wonder whether there might be an analogous way to derive \emph{de Sitter} spacetime in at least some of the settings relevant to the dS/CFT correspondence \cite{Strominger:2001pn}, or more generally, of the various approaches to de Sitter holography, recently reviewed in \cite{Galante:2023uyf}. In the present note we will answer this question in the affirmative.

To orient ourselves, consider the well-known fact that de Sitter can be obtained from anti-de Sitter via double Wick rotation. Concretely, if in (\ref{adsmetric}) we replace 
$x_0\to i x_4$ and $x_4\to it$, we obtain the dS$_5\times$H$_5$ geometry
\begin{equation}
\begin{aligned}
ds^2&=\frac{\tau^2}{L^2}\left(dx_1^2+\ldots+dx_4^2\right)-\frac{L^2}{\tau^2}d\tau^2+L^2 d\mathrm{H}_5^2
\\
&=\frac{\tau^2}{L^2}\left(dx_1^2+\ldots+dx_4^2\right)+\frac{L^2}{\tau^2}\left(-dt^2+dx_5^2+\ldots+dx_9^2\right)~,
\end{aligned}
\label{dsmetric}
\end{equation}
where $\tau^2\equiv t^2-x_5^2-\ldots-x_9^2$, and $d\mathrm{H}_5^2$ denotes the line element of a unit 5-dimensional hyperbolic space. The asymptotic behavior as $\tau\to\infty$ now enforces the scaling
\begin{equation}
ds^2=\omega\, 
\left(dx_1^2+\ldots+dx_4^2\right)
+\omega^{-1}
\left(-dt^2+x_5^2+\ldots+dx_9^2\right)
\label{cd3metricscaling}
\end{equation}
with $\omega \to \infty$. This displays two important differences with respect to (\ref{nrd3metricscaling}). The first is that time is now a transverse coordinate, indicating that (\ref{dsmetric}) describes a stack of branes with purely spacelike 4-dimensional worldvolume, i.e., an S3-brane \cite{Gutperle:2002ai}. The second difference is that, due to the positioning of $t$, the lightcone in the transverse directions is now unaltered, implying that physics in that sector is relativistic, whereas the lightcone in the longitudinal directions now closes up, identifying the $\omega\to\infty$ limit as the opposite of nonrelativistic, i.e., \emph{Carrollian} \cite{LevyLeblond:1965abc,SenGupta:1966qer}. 
This is an interesting connection, because Carrollian physics has been much explored in recent years, for various important applications. For a recent review, see \cite{Bagchi:2025vri}. More specifically, what we encounter is a $p$-brane generalization of the standard, particle Carrollian limit, henceforth abbreviated $p$C, along the lines of \cite{Bergshoeff:2023rkk,Cardona:2016ytk,Barducci:2018wuj,Roychowdhury:2019aoi,Bergshoeff:2020xhv,Kluson:2022jxh,Bidussi:2023rfs,Bagchi:2023cfp,Harksen:2024bnh,Casalbuoni:2024jmj,Bagchi:2024rje}, but differing in the property that time is not longitudinal to the brane, but transverse, as in \cite{Blair:2023noj}.

In \hfill the \hfill subsequent \hfill sections \hfill we \hfill will \hfill spell \hfill out \hfill this \hfill relation \hfill between \hfill Carrollian \\
S-brane theories and de Sitter holography. Our outline and main results are as follows.  

In Section~\ref{dpspsubsec}, we show that (\ref{dsmetric}) can be obtained physically by acting on (\ref{adsmetric})  with T-duality along directions $x_0,x_4$, rather than through a formal double Wick rotation.
This can be carried out before or after the near-horizon limit, and we choose to illustrate the former option, demonstrating how to get from the familiar asymptotically flat D3 black brane to an asymptotically flat SD3 black brane. The generalization to D$p\to$ SD$p$ immediately follows, and the outcome is given in (\ref{sdpbkg}). Our results agree with those of \cite{Hull:1998vg,Hull:1998fh}, where they were likewise obtained by T-duality. In a more complete check, we find that the metric, dilaton and Ramond-Ramond (RR) fields derived here are in precise agreement with one of the families of SD$p$-branes worked out in \cite{Bhattacharya:2003sh} by directly solving the supergravity equations of motion. 

In the latter paper, the family in question was obtained by imposing an `extremality condition' whose physical significance was not elucidated. In Section~\ref{noforcesubsec} we show that one manifestation of this extremal character is the fact that the backgrounds exert zero net force on a microscopic probe SD-brane of the same dimensionality and orientation as the SD black brane. This is in turn related to their half-BPS nature \cite{Hull:1998vg}.   
Importantly, the temporal T-duality 
flips the sign of the kinetic term for the RR fields,
signaling a passage from Type II to Type II$^{*}$ string theory \cite{Hull:1998vg,Hull:1998ym}. 
This reverses the sign of the associated energy-momentum tensor, transmuting the effective negative cosmological constant that sustains the original AdS background into a positive cosmological constant that enables our desired dS solution. It is due to this ghostlike kinetic term for the RR field that the no-go theorem of \cite{Maldacena:2000mw} can be evaded.\footnote{As emphasized in \cite{Hull:1998vg}, even if these ghost fields would be a clear sign of instability purely within the supergravity action, the fact remains that the complete Type II$^*$ string theory is equivalent to  Type II under temporal T-duality, and is therefore exactly as well- or ill-behaved as Type II on a timelike circle.} 

In Section~\ref{limitsubsec}, we turn our attention to the effect of the same T-duality operations on the various NR brane limits. As anticipated above, the resulting limits, given in (\ref{carrolllimit}) and written down previously in \cite{Blair:2023noj}, define Carrollian SD$p$-brane theories, henceforth \hfill abbreviated \hfill CSD$p$. The \hfill connection \hfill between \hfill NR \hfill and \hfill C \hfill via \hfill temporal \\ 
T-duality had been discussed previously in \cite{Blair:2023noj,Gomis:2023eav}.

Our final section, \ref{blackbranesubsec}, verifies that the near-horizon limit of the asymptotically flat SD$p$ black brane  (\ref{sdpbkg}) precisely coincides with the limit (\ref{carrolllimit}) that defines the CSD$p$ theory. The implication is that the near-horizon geometries (\ref{asymptoticallycarrollbkg}), which we show to be conformal to dS$_{p+2}\times$H$_{8-p}$, are precisely the SD$p$ black branes in CSD$p$ theory. They interpolate between an inner Lorentzian core and an outer  Carrollian asymptopia. The statement of open-closed string duality then equates Type II$^*$ string theory (or equivalently, due to the asymptotics, CSD$p$ theory) on these backgrounds to the worldvolume theory on a stack of microscopic SD$p$-branes, i.e., maximally supersymmetric $U(N)$ Yang-Mills (MSYM) theory on $p+1$ Euclidean dimensions.\footnote{In the terminology of \cite{Hull:1998vg}, we are referring here to  Euclidean rather than Euclideanized YM theory. It can be derived by starting with  standard Lorentzian $\mathcal{N}=1$ SYM  in $(9+1)$-Minkowski and applying dimensional reduction over time and $8-p$ spatial directions.\label{euclideanfoot}}    For $p=3$, this is precisely the dS/CFT correspondence proposed in \cite{Hull:1998vg} as a direct analogue of the classic AdS$_5\times$S$^5$ duality \cite{Maldacena:1997re}. The extension to  $p\neq 3$ yields (non)dS/(non)CFT equivalences directly analogous to the dualities in \cite{Itzhaki:1998dd}, which involve a conformally-rescaled version of AdS$_{p+2}\times$S$^{8-p}$ and call for different descriptions in different energy regimes. 

 In sum,  the results of  this note  provide  satisfactory  
 Carrollian/de Sitter counterparts to the nonrelativistic/anti-de Sitter findings of \cite{Blair:2024aqz,Guijosa:2025mwh}, with all of the brane theories in question being U-dual to one another. In complete parallel with \cite{Guijosa:2025mwh}, the lessons drawn in the previous paragraph can be synthesized into a sentence involving three boxed equations: we learn that the set of open-string/closed-string equivalences
\begin{equation}
\fbox{
\begin{minipage}{6cm}
\begin{center}
Type II$^*$ string theory on \\
stack of $N$ SD$p$-branes \\ 
in flat 10-dim Minkowski
\end{center}
\end{minipage} 
\huge
$=$ 
\normalsize
\begin{minipage}{6cm}
\begin{center}
Type II$^*$ string theory on\\
asymptotically flat \\
SD$p$ black brane
\end{center}
\end{minipage}}
\label{polchinski}
\end{equation}
yield in the low-energy/near-horizon limit the dS/CFT dualities\footnote{For simplicity, we employ the dS/CFT denomination even for $p\neq 3$, where the setup is in truth non-dS/non-CFT, and we write dS$_{p+2}\times$H$_{8-p}$ 
as a metonym for the near-horizon geometry of the SD$p$ black brane. In Section~\ref{blackbranesubsec} we show that the two are conformally related.}  
\begin{equation}
\fbox{
\begin{minipage}{6cm}
\begin{center}
$U(N)$ MSYM on \\
$(p+1)$-dim Euclidean space 
\end{center}
\end{minipage} 
\huge
$=$ 
\normalsize
\begin{minipage}{6cm}
\begin{center}
Type II$^{*}$ string theory on\\
dS$_{p+2}\times$H$_{8-p}$
\end{center}
\end{minipage}}~,
\label{hull}
\end{equation}
which generalize Hull's holographic proposal \cite{Hull:1998vg} beyond the particular case $p=3$, and can be reinterpreted as the statements of open-string/closed-string duality in Carrollian SD$p$ theory,\footnote{Upon recognizing that (\ref{hull}) is the same statement as (\ref{cpolchinski}), it becomes more natural to include in the field theory description the center-of-mass degrees of freedom for the SD$p$ stack, taking the gauge group in (\ref{hull}) to be $U(N)$ instead of just $SU(N)$. This point was first emphasized in \cite{Guijosa:2025mwh} for the NR D-brane theories.}
\begin{equation}
\fbox{
\begin{minipage}{6cm}
\begin{center}
CSD$p$ theory on \\
stack of $N$ SD$p$-branes \\ 
in flat 10-dim $p$C spacetime
\end{center}
\end{minipage} 
\huge
$=$ 
\normalsize
\begin{minipage}{6cm}
\begin{center}
CSD$p$ theory on\\
asymptotically flat $p$C \\
SD$p$ black brane
\end{center}
\end{minipage}}~.
\label{cpolchinski}
\end{equation} 
Again, the $p$C abbreviation refers to the $p$-brane Carrollian geometry that results from the scaling (\ref{carrolllimit}).  

All insights of \cite{Guijosa:2025mwh} carry over, including the fact that both sides of the correspondence (\ref{cpolchinski}) and therefore (\ref{hull}) are 10-dimensional and exhibit a mix of relativistic and Carrollian features, and the property that the flat $p$-Carrollian geometry is the preexisting substrate on top of which the Lorentzian dS$_{p+2}\times$H$_{8-p}$ spacetime is constructed by the entanglement of the Euclidean MSYM degrees of freedom. 
 
 To further widen our view, it is worth emphasizing that the duality web that includes the Type II$^*$ CSD$p$ and Type II NRD$p$  theories is only the more cleanly understood instance of what should be a total of four distinct webs of dual non-Lorentzian brane theories. This is so because from, say, the Type II perspective, there are two distinct decisions to make when applying the scaling (\ref{nrlimit}): whether it is the directions transverse ($\perp$) or longitudinal ($\,||\,$)  to the brane in question that will be non-Lorentzian (which will follow from time being chosen to be respectively longitudinal or transverse to said brane --- e.g., whether we are dealing with a D$p$ or an SD$p$  brane), and whether we scale $\omega\to\infty$ or $\omega\to 0$ (thereby applying a nominally nonrelativistic or Carrollian limit) to tear the Lorentzian metric apart. The four choices are thus
 \begin{enumerate}
     \item ($\perp ,\infty$): Type II NRD$p$ theories, dual to Type II$^*$ CSD$p$ theories.
      \item ($\perp ,0 $): Type II CD$p$ theories, dual to Type II$^*$ NRSD$p$ theories.
      \item ($\,||\, ,\infty$): Type II CSD$p$ theories, dual to Type II$^*$ NRD$p$ theories.
      \item ($\,||\, ,0 $): Type II NRSD$p$ theories, dual to Type II$^*$ CD$p$ theories.
 \end{enumerate}

In this paper we focus on option 1. The Type II black brane in option 2 can be easily seen to involve a purely Carrollian geometry, instead of an asymptotically Carrollian one. Depending on whether $N$ is held fixed or scaled to infinity, this Carrollian brane is flat or merely asymptotically flat. Options 3 and 4 are more challenging to understand because of the lack of supersymmetry, and because the parent black branes \cite{Bhattacharya:2003sh,Chen:2002yq,Kruczenski:2002ap,Roy:2002ik} are generically singular when $\tau=L$, and therefore do not seem to allow a standard near-horizon limit. 
Notice that even though options 1 and 2 would both yield Euclidean YM theories on the worldvolume of the Type II$^*$ SD$p$-branes, there would be a physical distinction due to the inherent Carrollian or nonrelativistic nature: space would be absolute in the former case, but not in the latter. Likewise for options 3 and 4 viewed from the Type II perspective.     
A different kind of extension is to examine the effect of each of these four limits on branes of type and/or orientation that differs from those of the brane that gives name to the limit, similar to what was done for option 1 in \cite{Blair:2024aqz,Danielsson:2000gi,Gomis:2000bd,Danielsson:2000mu,Harmark:2025ikv}. We leave the exploration of all of these extensions for future work.  

\section{Extremal S-branes}
\label{sbransec}

\subsection{From D\emph{p} in Type II to SD\emph{p} in Type II*}
\label{dpspsubsec}

In this subsection, we show how to obtain a family of SD$p$-brane supergravity solutions in Type II* theories, starting from the well-known extremal D$p$-brane solutions in Type II. The Type IIA and IIB theories are respectively connected to Type IIB$^*$ and IIA$^*$ via T-duality along a timelike direction \cite{Hull:1998vg}. Moreover, just as in the case of \hfill the \hfill Type \hfill II \hfill theories, \hfill the \hfill two \hfill Type II* \hfill theories \hfill are \hfill related \hfill to \hfill one \hfill another \hfill by 
\\
T-duality along a spacelike direction. Throughout this section we will work with the full, asymptotically flat solutions, prior to taking any non-Lorentzian limit. 

For illustrative purposes, we will work out the explicit example of $p = 3$. The generalization to arbitrary $p$ is then straightforward. We begin with the extremal D3 black brane solution in Type IIB supergravity\footnote{As in \cite{Guijosa:2025mwh}, we purposefully shift ${C}_{0123}$ by a constant, so that it asymptotes to $ g_s^{-1}$ instead of vanishing at infinity. This is convenient \cite{Danielsson:2000gi,Gomis:2000bd,Danielsson:2000mu} for the NRD3 and related limits that will be applied in Section \ref{carrollsec}. To avoid clutter we leave implicit the non-longitudinal components of $ C_{(4)}$, which ensure that $ F_{(5)}=d C_{(4)}$ is self-dual. Also, here and in what follows we let $\phi$ refer only to the fluctuation of the dilaton field on top of its asymptotic value $\phi_0\equiv \ln g_s$.} \cite{Horowitz:1991cd},
\begin{equation}
\begin{aligned}
d s^2 &= H( r)^{-1/2}\left(-d t^{\,2}+\ldots + d x_3^{\,2}\right)\; + H( r)^{1/2}\left(d x_4^{\,2}+\ldots + d x_9^{\,2} \right)~, 
\\[0.5em]
 C_{(4)} &= g_s^{-1} H( r)^{-1}\, d t \wedge d x_1 \wedge d x_2 \wedge d x_3 ,\quad e^{ 2\phi}= 1 \:\:,
\\[0.5em]
H( r) &= 1 + \frac{ L^4}{ r^{\,4}}\,,\qquad  r^2 \equiv x_4^{\,2}+\ldots +x_9^{\,2}\,,\qquad  L^4 \equiv 4\pi  N  g_s  \ell_s^{\,4}\,.
\end{aligned}
\label{d3bkg}
\end{equation}
To this we wish to apply T-duality along $x_4$ and $t$. The spatial T-duality is standard, and we know its end result ahead of time, but it will be useful to go over it in detail, as a benchmark for the second, temporal transformation that we also aim to perform.    

As usual, T-duality is defined along an isometry direction, which requires that all background fields (metric, dilaton, RR-forms)  be independent of the coordinate we intend to T-dualize.  The fields of the extremal threebrane  are controlled by the harmonic function $H( r)$, and we can separate
\begin{equation}
 r^{\,2} \;=\;  x_4^{\,2} +  r^{\prime\,2}~, \qquad  r^{\prime\,2} \equiv x_5^{\,2}+\ldots +x_9^{\,2}~.
\end{equation}
So before T-dualizing along 
$ x_4$, we must get rid of the $ x_4$-dependence in $H$ by smearing out the D3-branes uniformly along that direction, writing 
\begin{equation}
H^{\prime}({r}^{\prime}, {x}_4) = 1 +  \sum_{k=-\infty}^\infty \frac{L^4}{\left(({x}_4 + k \ell)^2 + r^{\prime\,2} \right)^2}
\end{equation}
and taking the \(\ell \to 0\) limit with $ N/\ell$ fixed, and therefore ${L}^4/\ell$  fixed.\footnote{Physically, scaling $ N\propto \ell \to 0$ ensures that the D3-brane charge density per unit length remains finite, as it should: rather than placing stacks of $N$ branes each in an infinitely dense periodic array, we are completely delocalizing each of a finite number of branes.} The sum can then be approximated by the integral
${\ell^{-1}} \int_{-\infty}^\infty dy$, with $y \equiv {x}_4 + n\ell$,
implying that
\begin{equation}
 H^{\prime}( r^{\prime})\;=\;1+\frac{ L^4}{\ell}\int_{-\infty}^\infty\frac{dy}{(y^2+ r^{\prime\,2})^2}
=1+\frac{L^{\prime\,3}}{ r^{\prime\,3}}~,
\label{H for D4}
\end{equation}
where $ L^{\prime\,3} \equiv \pi L^4/2\ell=2\pi^2 N g_s l_s^4/\ell$. 

Now we are in an appropriate position to perform T-duality along $ x_4$, applying the standard Buscher rules  \cite{Buscher:1987sk,Buscher:1987qj,Bergshoeff:1995as,Myers:1999ps}
\begin{equation}
\begin{aligned}
g'_{44}
&= \frac{1}{g_{44}}
= H^{\prime\, -1/2}, 
 \\
e^{\phi'} 
&= e^{\phi}/\sqrt{g_{44}}
= H^{\prime\, -1/4}~,\\
g_s^{\prime}C'_{01234}
& = g_s C_{0123}= H^{\prime\, -1}~,
\end{aligned}
\label{buscher}
\end{equation}
with the new asymptotic coupling and compactification radius given by
\begin{equation}
g_{s}^{\prime} = g_s\frac{l_s}{R}~,\quad R^{\prime} = \frac{{l}_s^2}{R} ~.
\label{gR}
\end{equation}
Naturally, we obtain the D4 black brane solution in Type IIA,
\begin{equation}
\begin{aligned}
ds^{\prime\, 2}
&= H^{\prime\,-1/2}\Bigl(-\,d t^2 + d x_1^{\,2} +\ldots d x_4^{\prime\,2}\Bigr)
+\; H^{\prime\, 1/2}\Bigl(d r^{\prime\,2} +  r^{\prime\,2}\,d\Omega_4^2\Bigr)~,\\[0.4em]
e^{\phi'} 
&=  H^{\prime\,-1/4}~,
\qquad
C'_{(5)} = g_s^{\prime\,-1}H^{\prime\, -1}\,d t\wedge d x_1\wedge d x_2\wedge d x_3\wedge d x_4^{\prime}~.
\end{aligned}
\label{d4bkg}
\end{equation}
Charge quantization implies that $L^{\prime\,3}=\pi N^{\prime}g^{\prime}_s l_s^3$, from which we read off the relation $N^{\prime}=2\pi NR/\ell$. 

After this reminder of the familiar spatial T-duality, we now turn to the somewhat unfamiliar application of temporal T-duality \cite{Moore:1993zc}, to obtain the extremal SD$3$ black brane solution of type IIB$^{*}$, which was first discussed in \cite{Hull:1998vg}, under the name of E4-brane. It was shown in \cite{Hull:1998ym} that T-duality along time involves the \emph{same} Buscher rules as above. In our context, they read
\begin{equation}
\begin{aligned}
g^{\prime\prime}_{ t t}
&= \frac{1}{g^{\prime}_{ t t}}
= - H^{\prime\, 1/2}~, 
 \\
   e^{\phi''}&=e^{\phi'}/\sqrt{-g^{\prime}_{ t t}}
= 1~, 
\\
g^{\prime\prime}_s C^{\prime\prime}_{1234}&= g^{\prime}_s C^{\prime}_{01234}= H^{\prime\, -1}~,
\end{aligned}
\label{buschert}
\end{equation}
with couplings and compactification radii related through 
\begin{equation}
g_{s}^{\prime\prime} = g^{\prime}_s\frac{l_s}{T^{\prime}}~,
\quad T^{\prime\prime} = \frac{{l}_s^2}{T^{\prime}} ~.
\label{gT}
\end{equation} 

Using all of these relations, the background (\ref{d4bkg}) becomes
\begin{equation}
\begin{aligned}
ds^{\prime\prime\, 2}
&= H^{\prime\,-1/2}\Bigl( d x_1^2 + d x_2^2 + d x_3^2 + d x_4^{\prime\, 2}\Bigr)
+\; H^{\prime\, 1/2}\Bigl(-d t^{\prime\prime\, 2}+d r^{\prime\, 2} +  r^{\prime\, 2}\,d\Omega_4^2\Bigr)~,
\\[0.4em]
e^{\phi''} 
&= 1\,,\qquad
C^{\prime\prime}_{(4)} = g_s^{\prime\prime\,-1} H^{\prime\,-1}\, d x_1\wedge d x_2\wedge d x_3\wedge d x^{\prime}_4~.
\end{aligned}
\end{equation} 
As expected, due to (\ref{H for D4}) the brane here is completely delocalized along the dual time direction $ t^{\prime\prime}$. We are interested in localizing it by applying the reverse of the smearing process discussed above, which entails replacing  
\begin{equation}
   H^{\prime}(r^{\prime})\quad\longrightarrow \quad H^{\prime\prime}( t^{\prime\prime}, r^{\prime})=1+\frac{L^{\prime\prime\,4}}{( t^{\prime\prime\,2}- r^{\prime\,2})^2}~,
\end{equation}
where $L^{\prime\prime\,4}\equiv 2\ell L^{\prime\,3}/\pi$.
As before, by matching the numerical constants we find that  $N^{\prime\prime}= N^{\prime}\ell/2\pi T^{\prime}$.

Altogether then, the background corresponding to a localized SD3 black brane is found to be
\begin{equation}
\begin{aligned}
ds^{\prime\prime\,2}&=  H^{\prime\prime\,-1/2}\Bigl(  d x_1^2 + \ldots +d x_4^2\Bigr)
+ H^{\prime\prime\, 1/2}\left(-d\tau^2+\tau^2d\mbox{H}_5\right)~,
\\
e^{2\phi''} 
&=1\,,\qquad
C^{\prime\prime}_{(4)} = g_s^{\prime\prime\,-1} H^{\prime\prime\,-1}\, d x_1\wedge d x_2\wedge d x_3\wedge d x^{\prime}_4~,
\\
H^{\prime\prime}&=1+\frac{L^{\prime\prime\,4}}{\tau^4}~,
\qquad
L^{\prime\prime\,4}\equiv 4\pi N^{\prime\prime}g_s^{\prime\prime}l_s^4~,
\end{aligned}
\label{sd3bkg}
\end{equation}
where $\tau^2 \equiv t^{\prime\prime\,2} -  r^{\prime\,2}$, $d$H$_5^2 \equiv d\rho^2 + \sinh^2(\rho)d\Omega_4^2$ and
$\rho\equiv\tanh\left(\frac{r^{\prime}}{ t^{\prime\prime}}\right)$.

The metric in (\ref{sd3bkg}) was first obtained in \cite{Hull:1998vg}, also via temporal T-duality. As promised, its near-horizon region is the dS$_5\times$H$_5$ geometry (\ref{dsmetric}), which was obtained from (\ref{adsmetric}) via double Wick rotation. We will return to this point in Section~\ref{carrollsec}. 
The complete set of fields in (\ref{sd3bkg}) matches exactly with the background obtained previously in \cite{Bhattacharya:2003sh} by solving the supergravity equations of motion\footnote{The authors of \cite{Bhattacharya:2003sh} initially set out to solve the equations of motion of Type II supergravity, so in their convention the RR potential ended up being purely imaginary, reflecting the negative-sign RR kinetic term that is a salient feature of Type II$^*$ supergravity \cite{Hull:1998vg}.} using a time-dependent ansatz and imposing a certain `extremality condition'.

The same procedure can be applied starting from a D$p$ black brane with arbitrary $p$ and performing T-duality along a transverse spatial direction followed by a T-duality in time. Equivalently,  we can  start with  a  D$(p+1)$  black  brane  and  just  perform 
T-duality along time. The resulting SD$p$ black brane solution takes the form (dropping all primes from now on)
\begin{equation}
\begin{aligned}
ds^2 &= H^{-1/2}\left(d x_1^2+\ldots+dx_{p+1}^2\right)+ H^{1/2}\left(-d\tau^2+\tau^2 d\mbox{H}_{8-p}^2\right)~,
\\[0.5em]
 C_{(p+1)} &= g_s^{-1}H^{-1} d x_1\wedge  \ldots\wedge d x^{p+1}~, 
 \quad 
 e^{\phi}=H^{(3-p)/4}
 \\[0.5em]
H &\equiv 1 + \frac{ L^{7-p}}{\tau^{7-p}}~, \quad L^{7-p}\equiv   (4\pi)^{(5-p)/2}\Gamma\left(\frac{7-p}{2}\right)N g_s \,l_s^{7-p}~.
\end{aligned}
\label{sdpbkg}
\end{equation}
These backgrounds again match perfectly with \cite{Hull:1998fh}, and with the family of `extremal' solutions obtained in \cite{Bhattacharya:2003sh}. 

In the following subsection we will provide a physical interpretation to the `extremality condition' of \cite{Bhattacharya:2003sh}, by showing that the time-dependent background (\ref{sdpbkg}) satisfies a no-force condition for the corresponding SD$p$-brane probe, as expected from an extremal solution that is assembled from an unexcited collection of such probes. This extremality property is in turn related to the fact that (\ref{sdpbkg}) is a BPS configuration \cite{Hull:1998vg}.

\subsection{No-force condition for extremal  SD$p$-branes}\label{noforcesubsec}

In the standard story of D$p$-branes in Type II, the extremal solution corresponds to a configuration that saturates the BPS bound, implying that the RR charge of the brane is equal to its mass (in appropriate units). This saturation leads to the preservation of a fraction of the supersymmetries ($1/2$ for the simplest D-brane systems) and ensures the existence of a no-force condition, meaning that the attractive gravitational and dilatonic interactions exactly cancel against the repulsive RR force. As a result, if one places an identically oriented probe D$p$-brane on the extremal supergravity background, the probe experiences no net force, a fact that can be explicitly verified by analyzing the worldvolume action of the probe. 

Interestingly, a similar notion of extremality can be discussed in the context of some spacelike branes. The corresponding supergravity backgrounds are generally non-supersymmetric due to their inherent time dependence \cite{Gutperle:2002ai,Chen:2002yq,Kruczenski:2002ap}, but in the context of Type II$^*$, the configurations  obtained from D$p$-branes in Type II via timelike T-duality do preserve supersymmetry \cite{Hull:1998vg}. This again points towards equality between mass and charge, in a liberal generalization of these concepts. `Charge' now refers \cite{Gutperle:2002ai} to a quantity obtained by integration of flux along a Gaussian surface that extends in time to completely surround the S-brane. This quantity is therefore \emph{not} conserved along time, but along a spacelike direction. In the present subsection, we show that such equality again leads to a no-force condition for specific types of probes. 

To see this explicitly, we will follow the same procedure as in \cite{Tseytlin:1996hi}, but particularized to the extremal SD$p$-brane solution given in (\ref{sdpbkg}). Specifically, we consider a probe SD$p$-brane placed on the background of the extremal SD$p$ black brane. The dynamics of the probe is governed by the Dirac-Born-Infeld  action supplemented by the Chern-Simons term:
\begin{equation}
S_{\mbox{\tiny SD}p}
= -\,T_{\mbox{\tiny SD}p} \int d\sigma^{p+1}\left[e^{-\phi}\sqrt{\det g_{ab}}
  - \frac{1}{(p+1)!}\epsilon^{a_1...a_{p+1}}c_{a_1...a_{p+1}} \right]~,
  \label{Sp action}
\end{equation}
where
we choose the static gauge $\sigma^a = x^a$ for $a = 1, \dots, p+1$, and $g$ and $c$ are the pullbacks of the corresponding background fields,
\begin{equation}
\begin{aligned}
    g_{ab} &\equiv G_{\mu\nu}\,\partial_a X^\mu \partial_b X^\nu = G_{ab} + G_{\alpha\beta}\,\partial_a X^{\alpha} \partial_b X^{\beta}~, 
    \\
    c_{a_1\cdots a_{p+1}} &\equiv C_{\mu_1\cdots \mu_{p+1}}\,\partial_{a_1} X^{\mu_1} \cdots \partial_{a_{p+1}} X^{\mu_{p+1}} =  C_{a_1\cdots a_{p+1}} + \cdots. \label{induced}
\end{aligned}
\end{equation}
Expanding the action \eqref{Sp action} using (\ref{induced}), we can isolate an effective potential from the part that depends on derivatives of the transverse embedding fields $X^{\alpha}$,
\begin{equation}
S_{\mbox{\tiny SD}p} = -T_{\mbox{\tiny SD}p} \int d^{p+1}x \left[ V_{\rm eff} + \cdots \right], \label{Sp_action_reduced}
\end{equation}
where
\begin{equation}
V_{\rm eff}(\tau) = e^{-\phi(\tau)} \sqrt{\det G_{ab}(\tau)} -  C_{1\cdots(p+1)}(\tau)~.
\label{veff}
\end{equation}
We consider a configuration in which the probe is oriented parallel to the source and located at a fixed transverse position $X^{\alpha} = c^{\alpha}$ (with $\alpha = 0, p+2, \dots, 9$). Substituting the background fields (\ref{sdpbkg}) into (\ref{veff}), we find
\begin{equation}
    V_{\rm eff}(\tau) = \left( g_s^{-1} H(\tau)^{\tfrac{p-3}{4}} \right) \left( H(\tau)^{-\tfrac{p+1}{4}} \right) - g_s^{-1} H^{-1}(\tau)
= 0~.
\end{equation}
This confirms that the probe feels no net force, in agreement with the interpretation of the background as an extremal configuration.

\section{Carrollian Brane Theories}
\label{carrollsec}

In this section we will discuss non-Lorentzian limits. In consonance with the notation of frequent use in the recent literature on NR brane physics, hatted variables will refer to `parent' quantities (i.e., those in the original, pre-limit theories), and unhatted variables will denote `daughter' quantities (i.e., those that are physically relevant after applying a non-Lorentzian limit). 

\subsection{The defining limit for Carrollian SD\emph{p}-brane theories}
\label{limitsubsec}

Knowing from Section~\ref{dpspsubsec} that temporal T-duality converts a D$(p+1)$-brane to an SD$p$-brane \cite{Hull:1998vg}, we will first review the limit that gives rise to the nonrelativistic D$(p+1)$-brane (NRD$(p+1)$) theory \cite{Danielsson:2000gi,Gomis:2000bd}. 

We begin with Type IIA/B string theory with string coupling $\hat{g}_s$ and string length $\hat l_s$ on 10-dimensional Minkowski space, equipped with a foliation that distinguishes between $p+2$ longitudinal and $8 - p$ transverse directions. For convenience \cite{Danielsson:2000gi}, we turn on a constant critical value for the RR $(p+2)$-form potential along the longitudinal directions, and compactify the spatial longitudinal directions on a $(p+1)$-torus, for simplicity taken to be rectangular and with radii set to a common value $\hat{R}$. The initial nonvanishing background fields are thus given by
\begin{equation}
 \begin{aligned}
    \hat{G}_{\mu \nu}& = \eta_{\mu \nu}~, \quad 
    \hat{C}_{(p+2)} = \hat{g}_s^{-1} d\hat{t} \wedge d\hat{x}_1 \wedge \cdots \wedge d\hat{x}_{p+1}~,
    \\ 
    \hat{x}_i& \simeq \hat{x}_i + 2\pi \hat{R} \quad (i = 0, \dots, p+1)~.
\end{aligned}
 \label{minkowski}
\end{equation}
For abritrary $p$, the nonrelativistic  D($p+1$)-brane (NRD$(p+1)$) theory is obtained by taking the limit \cite{Danielsson:2000gi,Gomis:2000bd}
\begin{equation}
\begin{aligned}
    \omega &\to \infty~, \quad \text{with} 
    ~\\
    \hat{x}^A &= \omega^{1/2} x^A \quad (A = 0, 1, \dots, p+1)~, 
    \\
    \hat{x}^{A'} &= \omega^{-1/2} x^{A'} \quad (A' = p+2, \dots, 9)~, 
    \\
    \hat{g}_s &= \omega^{\frac{p-2}{2}} g_s~, \quad \hat l_s = l_s~, 
    \\
    \hat{N} &= N~, \quad \hat{R} = \omega^{1/2} R~. 
\end{aligned}
\label{nrlimit}
\end{equation}
Applying this limit to the configuration (\ref{minkowski}), we can equivalently state that NRD$p$ theory is defined by scaling the background fields according to
\begin{equation}
\begin{aligned}
    d\hat{s}^{\,2} &= \omega  \, dx^A dx^B \eta_{AB} + \omega^{-1} \, dx^{A'} dx^{B'}\delta_{A'B'} ~, 
    \\
    \hat{C}_{(p+2)} &= \omega^2 g_s^{-1} \, dt \wedge dx_1 \wedge \cdots \wedge dx_{p+1}~, \\
    \hat{g}_s &= \omega^{\frac{p-2}{2}} g_s~.
\end{aligned}
\end{equation}
As reviewed in the Introduction, this tears the Lorentzian metric apart, giving rise to a $p$-brane Newton-Cartan ($p$NC) geometry \cite{Andringa:2012uz,Kluson:2017abm,Pereniguez:2019eoq,Blair:2021waq,Ebert:2021mfu,Novosad:2021tlq,Bergshoeff:2023rkk,Ebert:2023hba,Bergshoeff:2024ipq} describing a longitudinal subspace that is still Lorentzian and a transverse subspace that is nonrelativistic. The scaling also modifies the spectrum in such a way that longitudinal D$p$-branes become the lightest objects in the theory. 

Next, we consider T-duality along the time direction, which leads us to Type II$^*$ string theory. Using the Buscher rules (\ref{buschert}) we obtain
\begin{equation}
\begin{aligned}
    \hat{ds}^2 &= \omega \left( dx_1^2 + \cdots + dx_p^2 \right) + \omega^{-1} \left( -dt^2 + dx_{p+1}^2 + \cdots + dx_{9-p}^2 \right)~, 
    \\
    \hat{C}_{(p+1)} &= \omega^2 g_s^{-1} \, dx_1 \wedge \cdots \wedge dx_p~, 
    \\
    \hat{g}_s&= \omega^{\frac{p-3}{2}} g_s~.
\end{aligned}
\end{equation}
In parallel with (\ref{minkowski})-(\ref{nrlimit}), this can be equivalently expressed as starting from Type II$^*$ on the flat background with a  constant RR potential
\begin{equation}
    \hat{G}_{\mu \nu} = \eta_{\mu \nu}, \quad 
    \hat{C}_{(p+1)} = \hat{g}_s^{-1} \, d\hat x_1 \wedge \cdots \wedge d\hat x_p, \quad 
    \hat x_i \simeq \hat x_i + 2\pi \hat R \quad (i = 1, \dots, p).
    \label{minkowski-dual}
\end{equation}
and applying the rescaling
\begin{equation}
\begin{aligned}
    \omega& \to \infty~, \quad \text{with} 
    \\
    \hat{x}^{A}& = \omega^{1/2} x^{A}~, \quad A = 1, \dots, p~,
    \\
    \hat{x}^{A'}& = \omega^{-1/2} x^{A'}~, \quad A' = 0, p+1, \dots, 9~,  
    \\
    \hat{g}_s& = \omega^{\frac{p - 3}{2}} g_s~, \quad \hat{l}_s = l_s~, \quad \hat{N} = N~, \quad
    \hat R = \omega^{1/2}R~.
\end{aligned}
\label{carrolllimit}
\end{equation}
\emph{This procedure defines Carrollian SD$p$ brane (CSD$p$) theory} \cite{Blair:2023noj}, emerging as a consistent limit of Type II$^*$ string theory backgrounds with large RR potential and anisotropic scaling between the longitudinal and transverse directions. More precisely, the above expressions give rise to CSD$p$ on asymptotically \emph{flat} $p$-Carrollian geometries, and just like in the NR case, it is possible to introduce nontrivial longitudinal and transverse vielbeine to discuss curved Carrollian geometries as well \cite{Blair:2023noj}.

Starting from (\ref{carrolllimit}), one can apply further spatial T-dualities and/or S-duality to reach other Type II$^*$ Carrollian brane theories, such as CSF1 and CSNS5.  And as explained at the end of the Introduction, it is also natural to envision Carrollian D$p$, SD$p$, F1, SF1,  NS5 and SNS5 brane theories in Type II, which would be related via temporal T-duality to NR brane theories in Type II$^*$.

\subsection{SD\emph{p} black brane and de Sitter space}
\label{blackbranesubsec}

We now apply the Carrollian SD$p$-brane limit \eqref{carrolllimit} to the extremal SD$p$ black brane solution given in (\ref{sdpbkg}). To conform with the notation in the present section, we regard all quantities in (\ref{sdpbkg}) as hatted. Under the limit, the proper time and curvature radius scale as
\begin{equation}
    \hat{\tau}^2 = \omega^{-1} \tau^2, \quad \hat{L}^{7-p} = \omega^{(p-3)/2} L^{7-p}~,
\end{equation}
implying that the harmonic function simplifies to
\begin{equation}
    H(\tau)= \omega^{2}\frac{L^{7-p}}{\tau^{7-p}}~.
\end{equation}
The background that results from the Carrollian limit is thus
\begin{equation}
\begin{aligned}
{d\hat s}^{\,2} &=\left(\frac{\tau}{L} \right)^{\frac{7-p}{2}}\left ( dx_{1}^{2}+\dots+dx_{4}^{2} \right )
+\left(\frac{L}{\tau} \right)^{\frac{7-p}{2}}\left (-d\tau^2+\tau^2 dH_{5} \right )~,  
\\
\hat{C}_{(p+1)}&=g_s^{-1}\left(\frac{\tau}{L} \right)^{7-p} dx_{1}\wedge \dots \wedge dx_{p+1~,}
\\
e^{\hat\phi }&=\left(\frac{L}{\tau} \right)^{\frac{(7-p)(3-p)}{4}}~, \quad L^{7-p}\equiv   (4\pi)^{\frac{5-p}{2}}\,\Gamma\left(\frac{7-p}{2}\right)N g_s \,l_s^{7-p}~.
\end{aligned}
\label{asymptoticallycarrollbkg}
\end{equation}

The first feature that interests us here is that this background is precisely the same as what results from zooming onto the near-horizon region  \( \hat\tau \ll \hat L\) of the extremal SD$p$ black brane (\ref{sdpbkg}) via a Maldacena-like \cite{Maldacena:1997re,Itzhaki:1998dd} low-energy limit. The latter would normally be phrased in changing units as 
\begin{equation}
    \hat{l}_s=\omega^{-1}l_s\to 0 \quad\mbox{with}\quad
    \hat{N},\hat{g}_{\mbox{\tiny YM}}^{\,2}\equiv(2\pi)^{p-2} \hat{g}_s\hat{l}_s^{\, p-3},\hat{\tau}/\hat{l}_s^{\,2}\quad\mbox{fixed,}
    \label{malda}
\end{equation}
so to compare with our convention, where the string length is held fixed, one must examine dimensionless quantities such as 
$\hat{g}_s,\hat{x}^A/\hat{l}_s,\hat{\tau}/\hat{l}_s$.
Just like the NRD$p$ limit (\ref{nrlimit}) and the near-horizon limit of the D$p$ black brane have been found \cite{Blair:2024aqz} to be identical,  \emph{the CSD$p$ limit (\ref{carrolllimit}) and the near-horizon limit (\ref{malda}) of the SD$p$ black brane (\ref{sdpbkg}) fully coincide.}

A second interesting feature is that, as a result of its origin in the CSD$p$ limit, the asymptotics of the solution (\ref{asymptoticallycarrollbkg}) precisely match with the Carrollian scaling  (\ref{carrolllimit}), under the identification $\omega\leftrightarrow (\tau/L)^{(7-p)/2}$. In other words, the near-horizon fields (\ref{asymptoticallycarrollbkg}) describe a background that is \emph{asymptotically Carrollian} (specifically, asymptotically $p$-Carrollian). In direct parallel with the notion of asymptotically nonrelativistic backgrounds \cite{Avila:2023aey}, asymptotically  Carrollian backgrounds are solutions of the equations of motion of the parent relativistic theory that smoothly interpolate between an inner Lorentzian core and an outer Carrollian asymptopia. Due to these asymptotics, (\ref{asymptoticallycarrollbkg}) is understood to be the SD$p$ black brane in CSD$p$ theory. As in \cite{Blair:2024aqz,Guijosa:2025mwh,Avila:2023aey}, the solution's inner Lorentzian core arises from a U-dual version of NR string theory's $\lambda\bar{\lambda}$ deformation, itself an instance of the much studied $T\bar{T}$ deformation \cite{Blair:2020ops}. If this deformation were present with a totally arbitrary coefficient function, it would situate us squarely back within the parent Type II$^*$ theory. But importantly, the coefficient is such that the deformation switches off asymptotically, consistent with the identification of (\ref{asymptoticallycarrollbkg}) as a background in CSD$p$ theory. 

The third and final feature of central importance for our story in this paper is the contact between (\ref{asymptoticallycarrollbkg}) and de Sitter spacetime. As discovered in \cite{Hull:1998vg} and reviewed in the Introduction, for \( p = 3 \) the  metric in (\ref{asymptoticallycarrollbkg}) is directly identified as \( \mbox{dS}_5 \times \mbox{H}_5 \). We will now show that the connection is more general, because for arbitrary $p$, the metric is conformal to de Sitter space. 

To make this explicit,  
following the approach of \cite{Skenderis:1998dq} we construct the `dual frame' metric via the specific conformal transformation
\begin{equation}
d\tilde{s}^{\,2}\equiv\left(e^{\hat{\phi}}\right)^{\frac{2}{p-7}}d\hat{s}^{\,2}
= \left(\frac{\tau}{L}\right)^{\frac{3-p}{2}}d\hat{s}^2~,
\end{equation}
leading to
\begin{equation}
  d\tilde{s}^2 = \left(\frac{\tau}{L}\right)^{5-p}\left(dx_1^2+\ldots+dx_{p+1}^2\right)+  \left(\frac{L}{\tau}\right)^{2}\left(-d\tau^2+\tau^2 dH_{8-p}\right)~.
\end{equation}
Next we perform a coordinate transformation $\tau\to\mathcal{T}$, with
\begin{equation}
    \left(\frac{\mathcal{T}}{\mathcal{R}}\right)^2 = 
    \left(\frac{\tau}{L}\right)^{5-p}, \qquad 
    \mathcal{R} \equiv \frac{2}{5-p} L~.
\end{equation}
The metric then becomes
\begin{equation}
    d\tilde{s}^2=\frac{\mathcal{T}^{\,2}}{\mathcal{R}^2}\left(dx_1^2+\ldots+dx_{p+1}^2\right)-\frac{\mathcal{R}^2}{\mathcal{T}^{\,2}}d\mathcal{T}^{\,2}+\left(\frac{5-p}{2}\right)^2\mathcal{R}^2\,d\mbox{H}^2_{8-p}~.
\end{equation}
This is manifestly the metric of \( \mbox{dS}_{p+2} \times \mbox{H}_{8-p} \), showing that the near-horizon Carrollian limit of the extremal SD$p$ black brane in Type II$^*$ (or equivalently, the SD$p$ black brane in CSD$p$ theory) indeed yields a geometry conformal to de Sitter space.

In the `open string' description, it follows from the results of \cite{Danielsson:2000gi,Gomis:2000bd} via temporal T-duality that the Carrollian limit (\ref{carrolllimit}) singles out precisely SD$p$-branes as the lightest objects in the theory, with their dynamics being described by $U(N)$ maximally supersymmetric Yang-Mills in $(p+1)$ Euclidean dimensions,\footnote{See footnote~\ref{euclideanfoot}.} with coupling 
${g}_{\mbox{\tiny YM}}^{\,2}\equiv(2\pi)^{p-2} {g}_s{l}_s^{\, p-3}$. 
We arrive then at a generalization of Hull's dS/CFT duality \cite{Hull:1998fh}, just as advertised in the boxed equation (\ref{hull}), or equivalently, (\ref{cpolchinski}).

\section*{Note added}

While this paper was in preparation, we received \cite{Blair:2025nno}, which has substantial overlap with our results.

\section*{Acknowledgments}
Our work was partially supported by DGAPA-UNAM grant IN116823.

%%%%%%%%%%%%%%%%%%% BIBLIOGRAPHY %%%%%%%%%%%%%%%%%%%
%\clearpage
%\newpage

\bibliography{CarrolldS}

\bibliographystyle{./utphys}

%\end{multicols}

\end{document}